# Coherent Atomically-Thin Superlattices with Engineered Strain

Saien Xie[1,2], Lijie Tu[1,*], Yimo Han[1,*], Lujie Huang[3], Kibum Kang[2], Ka Un Lao[3], Preeti Poddar[2], David A. Muller[1,4], Robert A. DiStasio Jr.[3], Jiwoong Park [2,3,†]

[1] School of Applied and Engineering Physics, Cornell University, Ithaca, NY 14853, USA.

[2] Department of Chemistry, Institute for Molecular Engineering, and James Franck Institute, University of Chicago, Chicago, IL 60637, USA

[3] Department of Chemistry and Chemical Biology, Cornell University, Ithaca, NY 14853, USA.

[4] Kavli Institute at Cornell for Nanoscale Science, Cornell University, Ithaca, NY 14853, USA.

* These authors contributed equally to this work.

† Correspondence to: jwpark@uchicago.edu

**Abstract**:

Epitaxy forms the basis of modern electronics and optoelectronics. We report coherent atomically-thin superlattices, in which different transition metal dichalcogenide monolayers—despite large lattice mismatches—are repeated and integrated without dislocations. Grown by a novel omnidirectional epitaxy, these superlattices display fully-matched lattice constants across heterointerfaces while maintaining a surprisingly isotropic lattice structure and triangular symmetry. This strong epitaxial strain is precisely engineered *via* the nanoscale supercell dimensions, thereby enabling broad tuning of the optical properties and producing photoluminescence peak shifts as large as 250 meV. We present theoretical models to explain this coherent growth as well as the energetic interplay governing the flat-rippled configuration space in these strained monolayers. Such coherent superlattices provide novel building blocks with targeted functionalities at the atomically-thin monolayer limit.

Epitaxial structures with coherent heterointerfaces, in which lattices of dissimilar materials are matched without dislocations, enable advanced scientific and technological applications, including multiferroic oxides with engineered strain and symmetry (*1*, *2*), high-performance quantum cascade lasers (*3*), and high-efficiency light emitting diodes (*4*). Two-dimensional (2D) coherent heterostructures and superlattices (Fig. 1A and 1B) can enable the realization of these applications in materials with atomic thicknesses. They can further serve as ultrathin building blocks for advanced stacking and hetero-integration with other materials (*5–7*), providing unique opportunities that are not available to their 3D analogs. Realizing this goal would require the integration of various 2D materials whose properties can be tuned by the strain required for coherent lattice matching, as well as a method for precisely controlling the superlattice dimensions while maintaining lattice coherence over the entire structure. Monolayer transition metal dichalcogenides (TMDs), many of which share similar crystal structures, provide an ideal material platform with diverse electrical, optical (*8*, *9*), piezoelectric (*10*, *11*), and valley properties (*12*). However, recent studies on TMD heterostructure synthesis have shown only limited capabilities toward realizing coherent 2D superlattices (*13–21*).

Here, we report coherent monolayer TMD superlattices with precisely controlled supercell dimensions and lattice coherence maintained over the entire structure, which result in broad tuning of their optical properties. We use $WS_2$ and $WSe_2$ as the two main TMDs for our heterostructures and superlattices (Fig 1A inset). They have a significant lattice mismatch (Δ) of approximately 4% with $WSe_2$ having the larger lattice constant. Fig. 1C shows scanning electron microscope (SEM) images of three representative $WS_2$/$WSe_2$ heterostructures with controlled supercell dimensions, where dark (bright) regions correspond to $WS_2$ ($WSe_2$) monolayers. Every triangular unit of $WS_2$ and $WSe_2$ shows a highly-symmetric, equilateral triangular shape of uniform width, each defined by straight, parallel heterointerfaces, that can be directly controlled with nanoscale precision. These widths can be as narrow as 20 nm (Fig. 1C center) and periodically modulated to form superlattices with different dimensions, represented by the two widths {$d_{WS2}$, $d_{WSe2}$}. Two examples of superlattices are shown, one primarily comprised by $WS_2$ ($d_{WS2} \gg d_{WSe2}$; Fig. 1C left) and the other by $WSe_2$ ($d_{WS2} \ll d_{WSe2}$; Fig. 1C right). These crystalline TMD superlattices are synthesized using a modulated metal-organic chemical vapor deposition (MOCVD) (*22*) process under a constant growth environment (temperature, pressure, and flow rate), while accurately controlling the concentrations of the chalcogen precursors with time (see supplementary materials and Fig. S1). Our superlattices are grown at a slow growth rate near thermodynamic equilibrium and show only the most stable W-zigzag edges (Fig. S2) (*23*, *24*). Heterostructures consisting of different metal and chalcogen elements can also be synthesized with a similar level of control (see Fig. S1C for an example of a $WSe_2/MoS_2/WS_2$ heterostructure).

Our $WS_2/WSe_2$ superlattices maintain lattice coherence over the entire crystal (Fig. 2). First, the superlattices are free of misfit dislocations. Fig. 2A shows an annular dark-field scanning transmission electron microscope (ADF-STEM) image near a heterointerface (dashed line) between $WS_2$ (lower) and $WSe_2$ (upper). The ADF-STEM data taken from a larger area (Fig. 2B) shows continuous lines of atoms with no misfit dislocations near the heterointerface across ~160 unit cells (shown after the inverse fast Fourier transform (FFT)), whereas one dislocation is expected every 25 unit cells on average for incoherent heterointerfaces with $\Delta \approx 4\%$, strongly suggesting a coherent heterointerface in our superlattices. Second, our superlattices display lattice constants that are uniform over the entire structure. Fig. 2C shows selective area electron diffraction (SAED) data measured from a representative superlattice {50 nm, 40 nm} within a region with a single epitaxy direction (denoted by the arrow). This data displays a single-crystal-like pattern with sharp and isotropic diffraction spots. Their positions are used to measure the lattice constants along the directions parallel ($a_{//}$) or perpendicular ($a_\perp$) to the heterointerfaces (schematic, Fig 1B), as well as the lattice mismatch along each direction (*e.g.*, $\delta_{//} = 2|a_{//,1} - a_{//,2}|/(a_{//,1} + a_{//,2})$). Diffraction data corresponding to $a_{//}$ (circles in Fig. 2C, enlarged in Fig. 2D) show a single diffraction spot with no separation, confirming perfect lattice matching ($\delta_{//} = 0$). Diffraction data corresponding to $a_\perp$ (squares in Fig. 2C, enlarged in Fig. 2D) also show very similar lattice constants; while two spots are observed, each originating from the $WS_2$ and $WSe_2$ regions (see below), the mismatch $\delta_\perp = 1.2\%$ is much smaller than $\Delta$. In contrast, the same diffraction spots measured from an incoherent $WS_2/WSe_2$ heterostructure display a 4% concentric separation with $\delta_{//} = \delta_\perp = \Delta$ (Fig. 2E, see Fig. S3 for original SAED patterns).

Lattice coherence is directly confirmed with nanoscale resolution over the entire $WS_2/WSe_2$ superlattice. For this, we use our newly-developed electron microscope pixel array detector (EMPAD), which measures local diffraction maps pixel-by-pixel, providing structural information for imaging with nanoscale resolution (see supplementary materials) (*25*). Figs. 2F and 2G show three maps generated based on EMPAD data taken from another superlattice {75 nm, 60 nm}, each plotting $a_{//}$, $a_\perp$, and lattice rotation (with the orientations of $a_{//}$ and $a_\perp$ defined differently for regions α, β, and γ). The $a_{//}$ map (Fig. 2F left) shows little contrast between $WS_2$ and $WSe_2$, generating a single histogram peak as shown in Fig. 2H left (region α; see Fig. S4 for β and γ histograms). The $a_\perp$ map shows a small contrast between the $WS_2$ and $WSe_2$ regions, generating two peaks (Fig. 2H left) centered 0.4% below (corresponding to $WS_2$) and 0.8% above (corresponding to $WSe_2$) the $a_{//}$ peak, resulting in $\delta_\perp = 1.2\%$ as seen in Fig. 2D. Third, the lattice rotation map resolves only one dislocation clearly (arrow) within the entire superlattice (lateral size ~3.2 μm), suggesting the existence of a dislocation-free, coherent lattice everywhere, including the boundary regions between the α, β, and γ regions. In contrast, incoherent heterostructures show arrays of dislocations at heterointerfaces (Fig. S5).

Figure 2 shows that $\delta_{//} = 0$ everywhere, confirming coherent heterointerfaces in our superlattice. In addition, we note that the lattice isotropy and rotational symmetry are maintained over the entire superlattice. Our TEM and EMPAD data confirm 1) nearly identical and isotropic unit cell dimensions for both the $WS_2$ and $WSe_2$ regions, 2) that the lattice orientation in our EMPAD map (Fig. 2G) is highly uniform (standard deviation < 1 mrad), consistent with the observed sharp and isotropic TEM diffraction spots (Fig. 2C and 2D), 3) that the superlattice is triangular with straight edges and heterointerfaces. This is in sharp contrast to the lattice anisotropy expected from conventional unidirectional epitaxy, where $a_{//}$ is matched for the epilayers and $a_\perp$ is free from any constraints, causing the superlattice to have a different symmetry than that of the original crystal. Instead, our superlattice is grown with coherent omnidirectional epitaxy (see supplementary text and Fig. S6), where regions of different epitaxy directions (α, β, and γ) coherently connect with each other while maintaining the same symmetry of the original crystal.

The perfect symmetry in our coherent superlattices imposes an additional constraint that requires identical values of $a_\perp$ for both $WS_2$ and $WSe_2$. This is further illustrated in Fig. 2I: when a triangular $WSe_2$ unit is replaced by $WS_2$, the latter needs to expand by the same amount in all directions (*i.e.,* larger $a_{//}$ and $a_\perp$) in order to coherently bridge the inner and outer triangular $WSe_2$ units. This ideal picture changes in real superlattices with finite bulk and shear moduli values, where the final structure will minimize the total elastic strain energy. In this case, the lattice would deviate from having identical $a_\perp$ values for $WS_2$ and $WSe_2$, resulting in $0 \lesssim \delta_\perp < \Delta$, as seen from our data. These observations can also be quantitatively predicted by coarse-grained simulations of these superlattices that account for both bond and angle interactions on an appropriate footing (Fig. 2H right and 2J). In this regard, it is the inclusion of angular interactions in particular, which accounts for the shear stiffness inside the TMD superlattice and thereby introduces local frustration (analogous to the antiferromagnetic triangular-lattice Ising model), that is key to predicting coherent omnidirectional epitaxy across the entire lattice as well as a small but non-vanishing $\delta_\perp$ (see Fig. S7 and supplementary materials).

This lattice coherence also results in a tensile (compressive) strain within the WS$_2$ (WSe$_2$) region in our superlattices, the magnitude of which varies depending on the supercell dimensions. Fig. 3A illustrates such strain control. For example, a smaller $d_{WS_2}$ or larger $d_{WSe_2}$ (with a small ratio $\rho = d_{WS_2}/d_{WSe_2}$) increases the tensile strain in WS$_2$ and decreases the compressive strain in WSe$_2$ as it brings $a_{//}$ and $a_\perp$ closer to the intrinsic values for WSe$_2$. In addition, the band structure of both WS$_2$ and WSe$_2$ are sensitive to the applied strain—the size of the direct band gap decreases (increases) when subject to tensile (compressive) strain (*26–29*). This strain-dependent band structure allows for broad tuning of the optical properties by superlattice design. Such strain-engineered optical properties are demonstrated in our experiment. Fig. 3B shows the false-color SEM images of five representative WS$_2$ (blue)/WSe$_2$ (yellow) coherent superlattices I-V with different $\rho$ (dimensions plotted in Fig. 3A). The resulting photoluminescence (PL) spectra show two peaks, with one corresponding to WS$_2$ and the other to WSe$_2$ (Fig. 3C inset). However, the WS$_2$ peak is red shifted from the intrinsic peak energy of 1.97 eV by $\Delta_{WS_2}$ while the WSe$_2$ peak is blue shifted from the intrinsic value of 1.61 eV by $\Delta_{WSe_2}$. Fig. 3C compares the normalized WS$_2$ peaks measured from superlattices I-V (each extracted from the full PL spectra) to the intrinsic WS$_2$ peak (dashed curve). Superlattices with smaller $\rho$ clearly show a larger $\Delta_{WS_2}$, as large as 250 meV (see Fig. S8 for representative original PL spectra). Fig. 3D further plots $\Delta_{WS_2}$ vs. $\Delta_{WSe_2}$ for additional superlattices with different supercell dimensions.

These PL characteristics are consistent with the strain engineered by the superlattice design. The positive values for both $\Delta_{WS_2}$ and $\Delta_{WSe_2}$ confirm the tensile (compressive) strain in $WS_2$ ($WSe_2$). Their magnitudes show a negative correlation, which is consistent with their expected negatively correlated strain magnitude (Fig. 3A). The largest $\Delta_{WS_2}$ of 250 meV, corresponding to a 3.4% uniaxial strain or a 1.4% isotropic biaxial strain (*26*), is consistent with the large tensile strain expected from superlattice V with a small $\rho = 0.1$. Moreover, the PL image (Fig. 3E right; taken at 1.75 eV) confirms that the highly red-shifted $WS_2$ PL peak indeed originates from the strained $WS_2$ region (SEM image of a similarly grown sample shown in Fig. 3E left). In general, superlattices with supercell dimensions below the diffraction limit (Fig. 3F, left and middle) show uniform PL intensities at their respective peak energies over the entire structure, with a similar uniformity compared to intrinsic $WS_2$ (Fig. 3F, right).

Strained thin films are known to relax through out-of-plane deformations such as wrinkles and ripples, which makes these films non-flat and their edges curved (*30–32*). However, our ultra-thin superlattices maintain lattice coherence and symmetry, even though they are highly strained and their edges are under stress during growth, alternating between compressive and tensile stress. This can be explained by the strong van der Waals (vdW) interactions between the superlattice and the underlying growth substrate ($SiO_2$ in our experiment), which keep the 2D superlattice flat. Fig. 4A plots the theoretically calculated total energy ($E_{tot}$) per $WSe_2$ of a strained $WSe_2$ monolayer on $SiO_2$ as a function of the out-of-plane ripple height (*A*, measured from peak to valley; see schematic in Fig. 4B). $E_{tot}$ consists of the elastic strain energy ($E_{el}$, triangles), computed using a macroscopic elastic energy model (that accounts for both stretching and bending energy components in an ultra-thin film), and the interlayer vdW binding energy between the $WSe_2$ and $SiO_2$ ($E_{vdW}$, squares), computed using an all-atom quantum-mechanical vdW energy model (*33*) (see supplementary text and Fig. S9). While the rippled state ($A \approx 3$ nm) that relaxes the compressive strain is lowest in energy, the energetic profile shows another minimum at $A = 0$ nm, corresponding to the flat state. We note here that these two states have similar energies because the reduction in $E_{el}$ roughly equals the increase in $E_{vdW}$ for the rippled state. The rippled and flat states are separated by an energetic barrier (with an activation energy of approximately 23 meV per $WSe_2$), since the increase in $A$ in the regime $0 < A < 1$ nm rapidly destabilizes $E_{vdW}$ without significantly stabilizing $E_{el}$. Fig. 4A thus predicts that the attractive vdW force from the substrate keeps $WSe_2$ flat and that the transition from the flat to rippled state can only occur in the presence of a significant perturbation. As a result, these theoretical findings suggest that the synthesis conditions in our experiment, which maintain a constant growth environment with no strong perturbations, allow the superlattice to remain flat and the growth edge straight during growth.

We note that the superlattices reported herein are subjected to a cool-down process after growth, from a relatively high growth temperature (600 ºC) to room temperature. This may introduce a significant perturbation (*e.g.*, thermal expansion/contraction of the superlattice and $SiO_2$) and induce ripples in the $WSe_2$, which is exactly what we observe in our samples. The atomic force microscope (AFM) height image of a representative $WS_2/WSe_2$ superlattice (Fig. 4C) clearly shows out-of-plane ripples in $WSe_2$ (schematically illustrated in Fig. 4B). These ripples run continuously across the $WSe_2$ stripes only and are periodic along the heterointerfaces, as shown in the enlarged AFM image (Fig. 4D top). The peak-to-valley height ($A$) is approximately 1-2 nm (measured from the AFM profile shown in Fig 4D bottom). This is surprisingly close to the value of $A$ for the lowest energy state in Fig. 4A, despite the use of a simple energetic model and an idealized superlattice geometry. We also observe that the ripple wavelengths ($\lambda$) for superlattices with different $d_{WSe_2}$ remain relatively constant (close to 30 nm as shown in Fig. 4E), with little dependence on $d_{WSe2}$ over one order of magnitude (ranging from 20 - 320 nm). This suggests that the presence of WS$_2$/WSe$_2$ interfaces has a minimal effect on the energetics of the ripple formation in this regime, and that the constant compressive strain in WSe$_2$ (even up to $d_{WSe2}$ = 320 nm) is released through rippling. This also explains the smaller range of $\Delta_{WSe2}$ shown in Fig. 3D.

For superlattices with $d_{WSe2}$ > 320 nm, however, the periodic ripples are no longer continuous across the WSe$_2$ area (see Fig. S10). This indicates the presence of an alternative strain relaxation mechanism, including the formation of misfit dislocations and a coherent length of ~320 nm for our WS$_2$/WSe$_2$ superlattices. We note that this coherent length is significantly larger than the critical thickness of 2 nm for the Si/Ge system with a similar $\Delta$ ~ 4% (*34*), as well as the critical thickness for the WS$_2$/WSe$_2$ system estimated using the People-Bean model (below 20 nm, see supplementary materials and Fig. S11) (*35*). A full explanation for such a long coherent length would require a general theory optimized for 2D, which is currently lacking. However, we expect that our stable superlattice growth conditions and a larger energetic barrier for dislocation formation in 2D system may account for the long coherent length. For example, there are limited configurations of covalent bonding for dislocations in 2D systems and no screw dislocations. Our demonstration of fully coherent 2D superlattices not only presents a platform for studying novel omnidirectional epitaxy, but also opens up the possibility of generating structures with tunable functionalities by design, at the atomically-thin monolayer limit.

**Acknowledgements:** We thank Sidney Nagel and Thomas Witten for helpful discussions. This work was primarily supported by the Air Force Office of Scientific Research (FA9550-16-1-0031, FA9550-16-1-0347, FA2386-13-1-4118) and the National Science Foundation (NSF) through the Cornell Center for Materials Research with funding from the NSF MRSEC program (DMR-1120296) and the Platform for the Accelerated Realization, Analysis, and Discovery of Interface Materials (PARADIM; DMR-1539918). Additional funding was provided by the Samsung Advanced Institute of Technology. Material characterizations including electron microscopy were supported by the Cornell Center for Materials Research (NSF DMR-1120296) and the MRSEC Shared User Facilities at the University of Chicago (NSF DMR-1420709). L.T., K.U.L., and R.D. acknowledge partial support from Cornell University through start-up funding. This research used resources of the Argonne Leadership Computing Facility at Argonne National Laboratory, which is supported by the Office of Science of the U.S. Department of Energy under Contract No. DE-AC02- 06CH11357 and resources of the National Energy Research Scientific Computing Center, which is supported by the Office of Science of the U.S. Department of Energy under Contract No. DE-AC02-05CH11231.

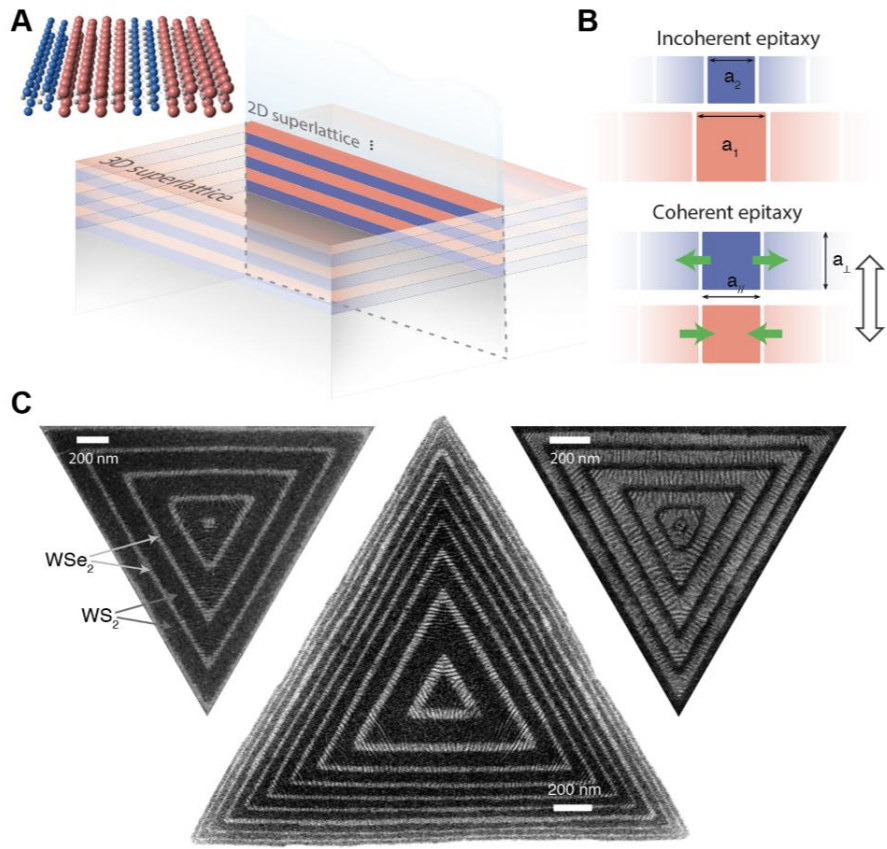

**Fig. 1. 2D monolayer TMD superlattices.** (**A**) Schematic of 3D and 2D superlattices. Inset: schematic of a lateral monolayer TMD superlattice. (**B**) Schematic of incoherent and coherent epitaxy, with the epitaxy direction represented by the outlined arrow. (**C**) SEM images of three monolayer $WS_2$/$WSe_2$ superlattices. Scale bars, 200 nm.

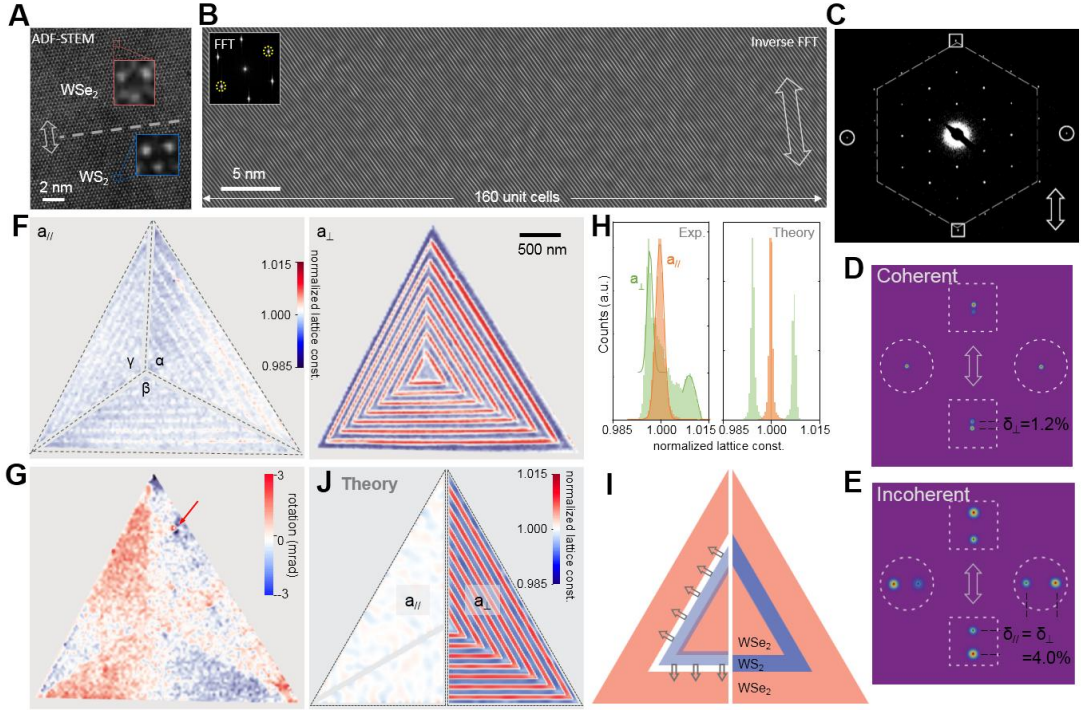

**Fig. 2. Lattice coherence of WS$_2$/WSe$_2$ superlattices.** (**A**) An ADF-STEM image at the heterointerface area between WS$_2$ and WSe$_2$ (epitaxy direction represented by the arrow, same for all). Scale bar, 2 nm. (**B**) Inverse FFT of an ADF-STEM image from a larger area near the heterointerface, based on the circled spots in its FFT (inset). Scale bar, 5 nm. (**C**) SAED pattern of superlattice {50 nm, 40 nm}, taken from an area of diameter of 280 nm. (**D**) Enlarged diffraction spots as indicated in (C). (**E**) The same diffraction spots as in (D) from an incoherent WS$_2$/WSe$_2$ heterostructure. (**F** and **G**) Spatial maps of normalized lattice constants $a_{//}$, $a_{\perp}$, and lattice rotation map of superlattice {75 nm, 60 nm}. Scale bar, 500 nm. (**H**) Histograms of $a_{//}$ and $a_{\perp}$ from experiment (for region α in (F)) and theory (for the superlattice depicted in (J)). (**I**) Schematic of the isotropic expansion of the WS$_2$ lattice in an omnidirectional coherent WS$_2$/WSe$_2$ heterostructure. (**J**) Composite maps of $a_{//}$ (left) and $a_{\perp}$ (right) of a superlattice with ratio $d_{WS_2}/d_{WSe_2}$ = 1.25, computed from a coarse-grained theoretical simulation (see Fig. S7 and supplementary materials).

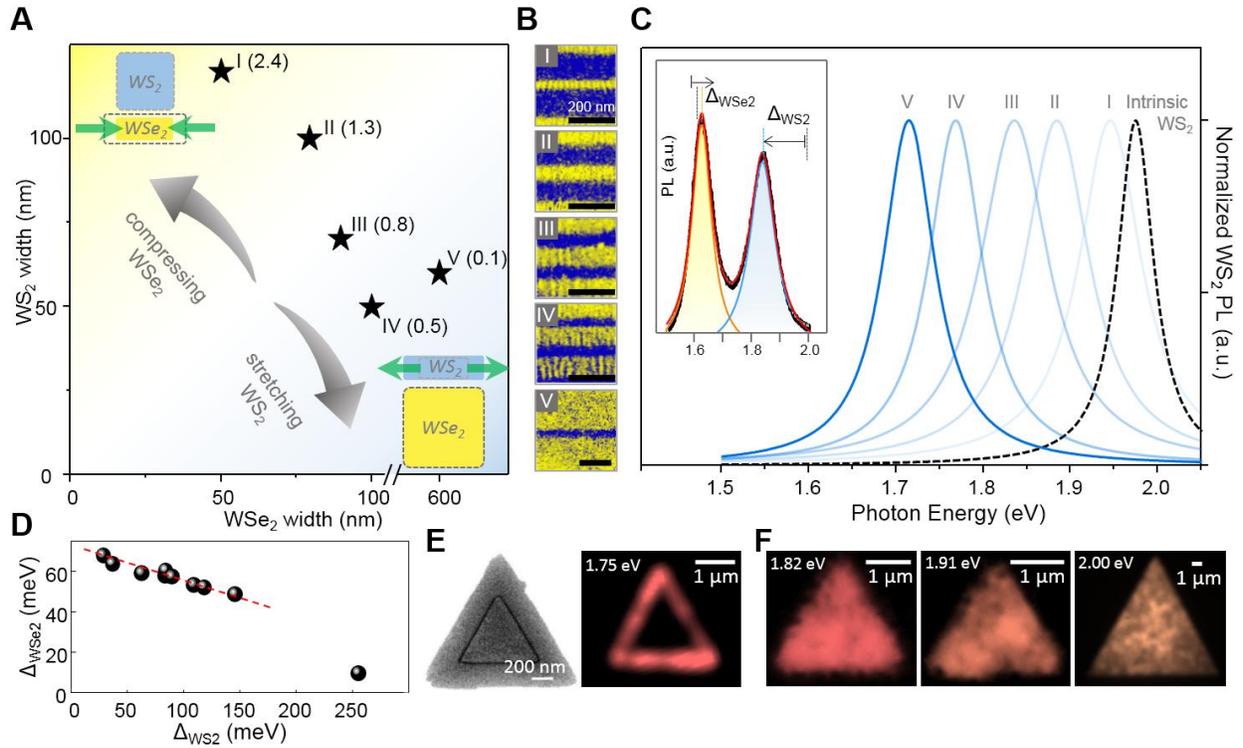

**Fig. 3. Strain engineering of PL of WS$_2$/WSe$_2$ superlattices.** (**A**) Plot of WS$_2$/WSe$_2$ superlattices I-V with different ratios ρ = d$_{WS2}$/d$_{WSe2}$ (values shown in parentheses). Insets: schematic of supercell dimension-dependent strain magnitude in the superlattice. (**B**) False-color SEM images of superlattices I-V. Scale bars, 200 nm. (**C**) Normalized PL spectra of WS$_2$ for intrinsic WS$_2$ (dashed line) and superlattices I-V. Inset: a representative PL spectrum of a WS$_2$/WSe$_2$ superlattice showing the WS$_2$ peak red shifted by Δ$_{WS2}$ and the WSe$_2$ peak blue shifted by Δ$_{WSe2}$. (**D**) Plot of Δ$_{WSe2}$-Δ$_{WS2}$ for WS$_2$/WSe$_2$ superlattices with different supercell dimensions. (**E**) Left: SEM image of a narrow WS$_2$ stripe embedded in WSe$_2$. Right: PL image of a heterostructure similar to the left, taken at photon energy of 1.75 eV. (**F**) PL images of two WS$_2$/WSe$_2$ superlattices at photon energies near their WS$_2$ peak positions (left and middle, at 1.82 eV and 1.91 eV, respectively) and an intrinsic monolayer WS$_2$ (right, at 2.00 eV). Scale bars, 1 μm.

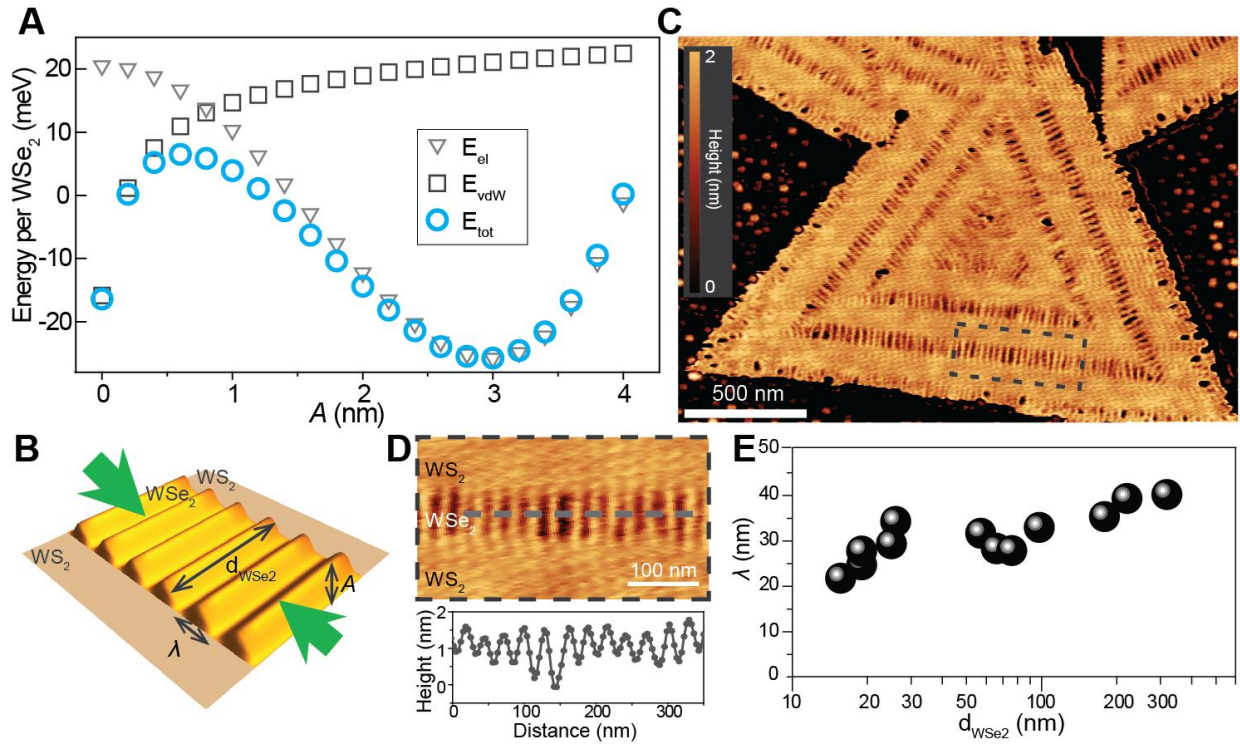

**Fig. 4. Formation of out-of-plane ripples in WSe₂.** (**A**) Theoretically calculated elastic strain energy ($E_{el}$), interlayer van der Waals binding energy ($E_{vdW}$), and total energy ($E_{tot}$) per WSe₂ as a function of WSe₂ ripple height (*A*) ($E_{el}$ and $E_{vdW}$ shifted by -32 meV and 53 meV, respectively, for clarity). (**B**) Schematic of rippled WSe₂ and flat WS₂, where the ripple wavelength (*λ*), *A*, and $d_{WSe2}$ are indicated. (**C**) AFM height image of a representative WS₂/WSe₂ superlattice. Scale bar, 500 nm. (**D**) Enlarged AFM image of the boxed area in (C) and height profile along the dashed line. Scale bar, 100 nm. (**E**) *λ* of superlattices with different $d_{WSe2}$.

# Supplementary Materials for

## Coherent Atomically-Thin Superlattices with Engineered Strain

Saien Xie, Lijie Tu, Yimo Han, Lujie Huang, Kibum Kang, Ka Un Lao, Preeti Poddar, David A. Muller, Robert A. DiStasio Jr., Jiwoong Park*

**Materials and Methods**

Modulated MOCVD growth of monolayer TMD superlattices

The synthesis of monolayer TMD superlattices was performed in a 2-inch quartz tube furnace (Fig. S1A). Tungsten hexacarbonyl (THC), molybdenum hexacarbonyl (MHC), diethyl sulfide (DES), and dimethyl selenide (DMSe) were selected as chemical precursors for W, Mo, S and Se, respectively. THC and MHC were kept in bubblers at a constant pressure of 800 Torr and introduced into the furnace with Ar as the carrier gas at room temperature. All precursors were introduced into the growth furnace with individual mass flow controllers (MFCs) regulating their flow rates. The growth of the superlattices was kept at a constant temperature of 600 °C and a total pressure of 2 Torr throughout. Flow rates of all precursors were kept constant during the growth except for the chalcogens: 20 sccm of THC or MHC with Ar as the carrier gas, 1 sccm $H_2$ and 350 sccm Ar, 0.3 sccm DES, and 0.4 sccm DMSe. The flow of the chalcogen precursors was alternated with breaks (typically 2 min) in between for purging purposes (see schematic in Fig. S1B). The typical growth time for each $WS_2$ or $WSe_2$ unit was a few minutes (denoted as $\{t_{WS2}, t_{WSe2}\}$), and the flow time could be controlled down to tens of seconds. For example, the superlattice shown in Fig. 1C left and right were grown by alternating {4-min $WS_2$, 3-min $WSe_2$} and {0.75-min $WS_2$, 7-min $WSe_2$}, respectively. The stable growth environment is crucial for coherent superlattice growth. In contrast, the synthesis with a temperature decrease (3 min, down to 300 °C) between different TMD growths result in heterostructures without lattice coherence. These heterostructures show incoherent diffraction patterns (Fig. 2E and Fig. S3 right), no PL tuning (Fig. S8 left), and no out-of-plane rippling (Fig. S10D). After the superlattice growth process, the furnace was naturally cooled down to room temperature in an inert environment with a constant flow of Ar. NaCl was placed at the upstream region of the furnace. The main growth substrate used was 285 nm $SiO_2$/Si. For more discussion of our MOCVD process, see ref. 22.

TEM analysis

*Sample preparation*: The monolayer TMD superlattice was spin-coated with PMMA A4, and then the substrate was etched in KOH 1M solution. After being rinsed in deionized water for three times, the PMMA supported superlattice was transferred to a TEM grid with 10 nm thick SiN windows, and the chip was annealed in an ultra-high vacuum ($10^{-7}$ Torr) at 350 °C for 3 hours to remove the PMMA.

*SAED and DF-TEM:* SAED patterns were taken using an FEI Tecnai T12 Spirit, operated at 80 kV. The selective area aperture has a diameter of 280 nm. The dark-field TEM images are taken by selecting specific diffraction spots (*36*).

*ADF-STEM*: ADF-STEM imaging was performed using a FEI TITAN operated at 120 kV with a ~15 pA probe current. A 30 mrad convergence angle and a ~40 mrad inner collection angle were used for all ADF-STEM images, whose contrast is proportional to $Z^\gamma$, where Z is the atomic number and $1.3 < \gamma < 2$.

*EMPAD*: Lattice spacing and rotation maps were performed using an electron microscope pixel array detector (EMPAD) (*25*), which is a high-speed, high dynamic range diffraction camera with the sensitivity to measure a diffraction pattern from a single atom (*37*). We used a 120-kV electron beam with a 0.5 mrad convergence angle to record diffraction patterns at every point in 2D real space scan to build up a 4-dimensional phase space map. By measuring the center of the second order diffraction spots using center of mass calculation, we mapped reciprocal lattice vectors for every scan position in real space. The inverse of these vectors provided the real-space lattice spacing maps shown in Fig. 2F (the lattice spacing in the armchair direction was multiplied by $\sqrt{3}$ to get the normalized $a_{//}$, therefore makes the unstrained hexagonal lattice have same $a_{//}$ and $a_{\perp}$). The rotation map in Fig. 2G was calculated by measuring the rotation angles of the reciprocal lattice vectors as a function of real-space scan positions. The color scale was normalized to the mean value of the entire sample region. (1 for the lattice constant map and 0 for the rotation map).

Photoluminescence measurements

The PL spectra were acquired using Horiba Lab RAM HR Evolution confocal Raman microscope, with a 532-nm laser excitation under ambient conditions. The PL images were taken with a widefield PL microscope using band pass filters with 10 nm bandwidth, and the center photon energies of 1.75 eV, 1.82 eV, 1.91 eV, and 2.00 eV, respectively (*38*).

**Supplementary Text**

Effective negative Poisson's effect in omnidirectional coherent epitaxy

Below, we explain the symmetry and lattice isotropy in the superlattice grown by omnidirectional coherent epitaxy, using a symmetry analysis. We use a square lattice to illustrate the symmetry constraint for simplicity. Conventional unidirectional coherent epitaxy only requires matched $a_{//}$ while $a_{\perp}$ is free of any constraints, thus $a_{\perp}$ of small (large) lattice becomes even smaller (larger), corresponding to $\delta_{\perp} > \Delta$ for positive Poisson's ratio (Fig. S6A). Therefore, the resulting heterostructure shows a different symmetry (rectangular) than that of the original crystal (square). However, in omnidirectional coherent epitaxy, the epilayers grown in all directions connect to each other without dislocation. As a result, the final heterostructure maintains the same symmetry with that of the original crystal, showing $\delta_\perp = 0$ (Fig. S6B). Similar results for our $WS_2$/$WSe_2$ superlattices is illustrated in Fig. 2I. The isotropic lattice constants result in an effectively negative Poisson's effect, regardless of the intrinsic Poisson's ratios of each component, and is universal for any coherent omnidirectional epitaxy system.

Coarse-grained simulations of the TMD superlattice

To study the equilibrium state of the entire coherent TMD superlattice shown in Fig. 2, we developed and employed a coarse-grained force-field model that accounts for nearest-neighbor bonding and angular interactions. In this approach, the atom types are "dressed W atoms" and can either be [$WS_2$] or [$WSe_2$] units centered at the corresponding W atomic position in the 2D superlattice (depicted as red or blue circles, respectively, in Fig. S7A and S7B). The total potential energy expression employed for the superlattice is $E_{latt} = \frac{1}{2}\sum_{bonds} k_b(r - r_0)^2 + \frac{1}{2}\sum_{angles} k_\theta(\theta - \theta_0)^2$, in which the first term represents a harmonic bond potential (with $k_b$ and $r_0$ representing the bond force constant and equilibrium bond distance, respectively) and the second term a harmonic angular potential (with $k_\theta$ and $\theta_0$ representing the angular force constant and equilibrium angle, respectively). Figs. S7A and S7B summarize the different types of bond and angle terms encountered in our energy expression for TMD superlattices.

In all of our simulations, the central (seed) triangle consists of [$WS_2$] atom types (with a side length of 20 dressed atoms). The superlattice is then constructed *via* the addition of alternating TMD bands, the next of which is 4 (or 8) layers of [$WSe_2$] atom types, followed by 5 (or 10) layers of [WS$_2$] atom types, etc. until the model superlattice consists of 14 total TMD bands (for direct comparison with the experimental superlattice considered in Fig. 2). In this initial configuration, all bond lengths were set to a value of $r$ = 3.22 Å, which is the mean of the experimental W—W distances in pure WS$_2$ ($r_0$ = 3.15 Å) and WSe$_2$ ($r_0$ = 3.29 Å) monolayers. For a given replica of the system (with a minimum of 6 replicas is always used) each of the initial positions of the dressed atoms are randomized (up to 5% of the initial bond distance) before $E_{latt}$ is minimized using second-order damped dynamics (with a damping parameter of $\gamma$ = 0.2 and fictitious time step of $\delta$t = 0.004, which have been optimized to ensure rapid convergence to the energetic minimum).

To parameterize the bond terms, we utilized the 2D Young's moduli for pure WS$_2$ ($Y_{2D}$ = 140 N/m) and WSe$_2$ ($Y_{2D}$ = 116 N/m) monolayers obtained from highly accurate density functional theory calculations (*39, 40*) to set $k_b$= 1.2 Å$^{-2}$ for [WS$_2$]—[WS$_2$] and $k_b$= 1.0 Å$^{-2}$ for [WSe$_2$]—[WSe$_2$]. For dressed atoms at the interface (in which [WSe$_2$]—[WS$_2$] or [WS$_2$]—[WSe$_2$]), the appropriate $k_b$ and $r_0$ are chosen to satisfy the outward growth mechanism of the TMD superlattice (see Fig. S7B for a graphical depiction).

Angular interactions were included to reflect the shear stiffness inside the TMD superlattice. In this regard, the larger the shear modulus is, the stronger is the tendency of the superlattice to maintain an equilateral triangular structure. Since the differential shear moduli for these TMD monolayers are unknown, we have studied a series of $k_\theta$ ranging from 0 rad$^{-2}$ (0 deg$^{-2}$, no angular interactions) to 40 rad$^{-2}$ (0.012 deg$^{-2}$, very strong angular interactions) with $\theta_0$ = $\pi$/3 rad = 60° in all cases. Quite importantly, we find that the inclusion of angular interactions is key to predicting the spatial patterns and histograms of both the parallel (a$_{//}$) and perpendicular (a$_\perp$) lattice constants (see Fig. S7C). With an intermediate-range value of $k_\theta$= 20 rad$^{-2}$ (0.006 deg$^{-2}$), we find excellent agreement with the experimental results for a$_{//}$ and a$_\perp$, as shown in Figs. 2F, 2H, and 2J.

Energetics governing the flat-rippled configuration space in strained $WSe_2$ monolayers

The model used to explore the energetics of the flat-rippled configuration space in a strained $WSe_2$ monolayer grown on an $\alpha$-$SiO_2$ substrate was assembled using the QuantumWise nanoscience simulation software package (see Fig. S9A). For consistency with the dimensions of the ripples observed from our TMD superlattices (see Fig. 4), we chose to model a $WSe_2$ monolayer ripple with a wavelength of $\lambda \approx 30$ nm and a width of $d_{WSe2} \approx 10$ nm (corresponding to the x- and y-axes, respectively, in Fig. S9A).

For the initial flat $WSe_2$ monolayer configuration, we started with an equilibrium $WSe_2$ monolayer consisting of 3,456 $WSe_2$ units with dimensions of x = 31.51 nm and y = 10.23 nm. Applying an isotropic axial strain of 4% in the x-direction (representing the lattice mismatch of $\Delta \approx 4\%$ between $WSe_2$ and $WS_2$ monolayers) coupled with transversal expansion in the y-direction (governed by Poisson's ratio of $\nu = 0.19$ for a $WSe_2$ monolayer (*39, 40*)), we arrived at our initial flat (and compressively strained) $WSe_2$ monolayer configuration with dimensions of x = 30.25 nm and y = 10.31 nm. We then proceeded to assemble the underlying $\alpha$-$SiO_2$ substrate (with an interlayer separation of $R = 3.5$ Å, reflecting a typical separation distance between van der Waals (vdW) heterostructures (*5*)) consisting of 42,336 $SiO_2$ units with dimensions of x = 30.27 nm and y = 10.32 nm (chosen to minimize the periodic mismatch between the $WSe_2$ monolayer and the substrate). In the z-direction, we utilized 6 $SiO_2$ layers to account for the bulk properties of the $SiO_2$ substrate. As a result, our model for exploring the flat-rippled configuration space in this system contains a total of 137,376 atoms (with 10,368 atoms in the $WSe_2$ monolayer and 127,008 atoms in the $\alpha$-$SiO_2$ substrate).

To introduce a ripple in the $WSe_2$ monolayer with height *A* (see Figs. 4B and S9A), we considered the series of (trough-to-trough) shapes formed by the sinusoidal family of functions ($\sin^n(x)$ with $n = 2, 4, 8$). In particular, these ripples were created by inducing a vertical displacement (*i.e.*, a z-coordinate change) in each atom contained in the WSe$_2$ monolayer based on its corresponding horizontal position (or x-coordinate) as follows:

(i) $\quad z = A\sin^2\left(\frac{\pi x}{\lambda}\right) = \frac{A}{2} + \frac{A}{2}\sin(\frac{2\pi x}{\lambda} - \frac{\pi}{2})$

(ii) $\quad z = A\sin^4\left(\frac{\pi x}{\lambda}\right) = A\sin^2(\frac{\pi x}{\lambda}) - \frac{A}{4}\sin^2(\frac{2\pi x}{\lambda})$

(iii) $\quad z = A\sin^8\left(\frac{\pi x}{\lambda}\right) = A\sin^4(\frac{\pi x}{\lambda}) - \frac{A}{2}\sin^2(\frac{\pi x}{\lambda})\sin^2(\frac{2\pi x}{\lambda}) + \frac{A}{16}\sin^4(\frac{2\pi x}{\lambda})$.

These sinusoidal transformations govern the shape (or profile) of the ripple and are more compactly represented by S$_1$, S$_2$, and S$_4$, respectively, throughout the text.

*Elastic energy for the WSe$_2$ monolayer*: The elastic energy (E$_{el}$) for the WSe$_2$ monolayer as a function of the ripple height, *A*, for ripple shapes generated using the sinusoidal transformations (defined above) was computed using continuum mechanics as a sum of the stretching (E$_s$) and bending (E$_b$) energy components (defined below), each of which is depicted in Fig. S9B. In this work, we adopted a harmonic potential energy expression (*i.e.*, Hooke's Law) to describe E$_s$ as follows: $E_s = \frac{1}{2} Y_{2D} d_{WSe2} \frac{(L-L_0)^2}{L_0}$, in which $Y_{2D}$ is the 2D Young's modulus (or 2D elastic stiffness) for a WSe$_2$ monolayer and $dL \equiv L - L_0$ is the amount by which the monolayer is stretched (or compressed) relative to the relaxed equilibrium length, $L_0$. Since the WSe$_2$ monolayer is comprised of only 3 atomistic layers, E$_s$ for this system is more appropriately described by $Y_{2D}$ instead of the 3D Young's modulus ($Y_{3D}$) as was found for other TMD systems such as MoS$_2$ as well as truly 2D materials such as graphene (*41, 42*). Throughout this work, we utilize the value of $Y_{2D}$ = 116 N/m, which was obtained from highly accurate density functional theory calculations (*39, 40*).

For $E_b$ we employed the following energy expression based on Euler buckling theory (*43, 44*): $E_b = \frac{1}{2}Bd_{WSe2}\int dx z''(x)^2$, in which $B$ is the bending stiffness and $z''(x)$ is the second derivative of the ripple shape with respect to $x$ (the stretching/compression direction). We note here that significant controversy still exists in the literature regarding the use of classical shell/plate theory for $B$ in 2D materials (such as TMD monolayers) as this approach suffers from the large uncertainty necessarily present in any definition of "thickness" in such ultra-thin nanofilms (*41, 45–47*). Throughout this work, we completely sidestep this issue by utilizing a recent experimentally derived value of $B = 12.4$ eV (corresponding to $WSe_2$ with zigzag chirality (*48*)) and therefore avoid the use of any measure of "thickness" in our energy expressions.

Since the initial flat $WSe_2$ monolayer has been compressed by 4% (corresponding to the lattice mismatch with $WS_2$), this configuration starts with a relatively large $E_s \approx 50$ meV per $WSe_2$ while $E_b = 0$ meV (see Fig. S9B). As the ripple forms and $A$ increases, $L$ increases towards $L_0$ releasing the compression strain and steadily decreasing $E_s$; although this is accompanied by a simultaneous monotonic increase in $E_b$, the total $E_{el} = E_s + E_b$ is still dominated by $E_s$ over the entire range of $A$ before the energetic minimum for all ripple shapes considered herein. The location of the minimum in $E_{el}$ is therefore primarily dictated by the minimum in $E_s$, although $E_b$ can play a more substantive role and therefore shift the minimum to lower $A$ values for ripple shapes induced by higher-order (and therefore more perturbative) sinusoidal transformations. For $A$ values beyond the minimum, both $E_s$ and $E_b$ steadily increase as $L$ is now larger than $L_0$ (corresponding to a stretched monolayer) and the degree of bending is becoming increasingly more severe.

*Interlayer van der Waals (vdW) energy*: To complete our description of the energetics governing the flat-rippled configuration space, we also account for the non-bonded and long-range vdW interactions between the $WSe_2$ monolayer and the underlying $SiO_2$ substrate. Here we utilized an all-atom quantum mechanical vdW correction (the so-called D3 method (*33, 49*) with modified Becke-Johnson (BJ) damping (*50*)). In this approach, the chemical environment (*via* the corresponding coordination number) of each atom has been accounted for in both the pairwise two-body (2B) vdW interactions, $E_{vdW-2B} = -\frac{1}{2}\sum_{AB}\sum_{n=6,8} s_n \frac{C_n^{AB}}{r_{AB}^n} f_{d,n}(r_{AB})$, and the beyond-pairwise three-body (3B) vdW interactions, $E_{vdW-3B} = \frac{1}{6}\sum_{ABC} \frac{C_9^{ABC}(3\cos\theta_a\cos\theta_b\cos\theta_c+1)}{(r_{AB}r_{BC}r_{CA})^3} f_{d,(3)}(r_{ABC})$, *via* the isotropic *n*-th order dispersion coefficients ($C_6^{AB}, C_8^{AB}, C_9^{ABC}$). In these expressions, $r_{AB}$ is the distance between atoms *A* and *B*, $\theta_a, \theta_b, \theta_c$ are the interior angles formed by the *ABC* triangle, and the scaling factors were set to s$_6$ = 1.0 and s$_8$ = 0.3589 throughout.

All vdW calculations were performed using an in-house version of the DFT-D3 program that has been modified to account for periodic boundary conditions (in the x- and y-directions) in both the atomic coordination number determination as well as the 2B and 3B interlayer vdW energy computations. Further modifications of the program were required to correctly deal with the different intrinsic periodicities of the WSe$_2$ monolayer and SiO$_2$ substrate.

In the initial flat WSe$_2$ monolayer, the interlayer vdW binding energy is largest (most negative) with a value of E$_{vdW}$ ≈ -70 meV per WSe$_2$ as the monolayer in this configuration is in closest contact with the underlying substrate (see Fig. S9B). As the ripple forms and *A* increases, the magnitudes of both the attractive E$_{vdW-2B}$ and repulsive E$_{vdW-3B}$ terms decrease, becoming less attractive and less repulsive, respectively. Since |E$_{vdW-3B}$| is only about 26-30% of |E$_{vdW-2B}$| (see Fig. S9B), their sum $E_{vdW} = E_{vdW-2B} + E_{vdW-3B}$ still remains attractive, although it becomes less attractive with increasing *A*. As expected, the overall decay rate (as a function of *A*) in the interlayer vdW binding energy decreases for ripple shapes generated with increasingly more perturbative (higher-order) sinusoidal transformations, reflecting the fact that more of the TMD monolayer is in close contact with the substrate in such ripple profiles.

*Total energy analysis*: The combination of $E_{el}$ and $E_{vdW}$ provides our total energy expression ($E_{tot}$ = $E_{el}$ + $E_{vdW}$), which is plotted as a function of *A* for the aforementioned sinusoidal ripple shapes in Fig. S9B (wherein $E_{tot}$ was computed with respect to the flat monolayer configuration). Since the energetic contributions from $E_{el}$ and $E_{vdW}$ in this system are competitive in nature and similar in magnitude, their combination predicts a distinct energetic barrier separating the initial flat (and highly strained) configuration with the minimum-energy rippled configurations emerging in the range of $A = 2 - 4$ nm. Here we find that the rippled configurations with higher-order sinusoidal shapes are most stable, with a relative energy range that can be tuned with respect to the flat configuration. In this regard, the values of *A* predicted by this simple hybrid macroscopic-microscopic theoretical model are in relatively good agreement with the corresponding experimental measurements (see Fig. 4D).

Estimation of critical thickness

The critical thickness of $WS_2/WS_2$ epitaxial system based on People-Bean model (*35*) is given by

$$h_c \cong \frac{1-\nu}{1+\nu} \frac{1}{16\pi\sqrt{2}} \frac{b^2}{a} \frac{1}{\Delta^2} \ln[\frac{h_c}{b}]$$

Where a = 0.329 nm is the lattice constant of $WSe_2$. b is the length of Burgers vector, and we take b = a = 0.329 nm as it corresponds to the easiest formed misfit dislocation; ν is the Poisson's ratio of $WSe_2$. The numerical result of the above equation is plotted in Fig. S11, showing a critical thickness of 17 nm (ν = -0.2), 9.8 nm (ν = 0), and 5.4 nm (ν = 0.2) for $WS_2/WSe_2$ epitaxy system with Δ = 4%. We note that the People-Bean model is known to overestimate the critical thickness for Δ < 6.2% (*51*).

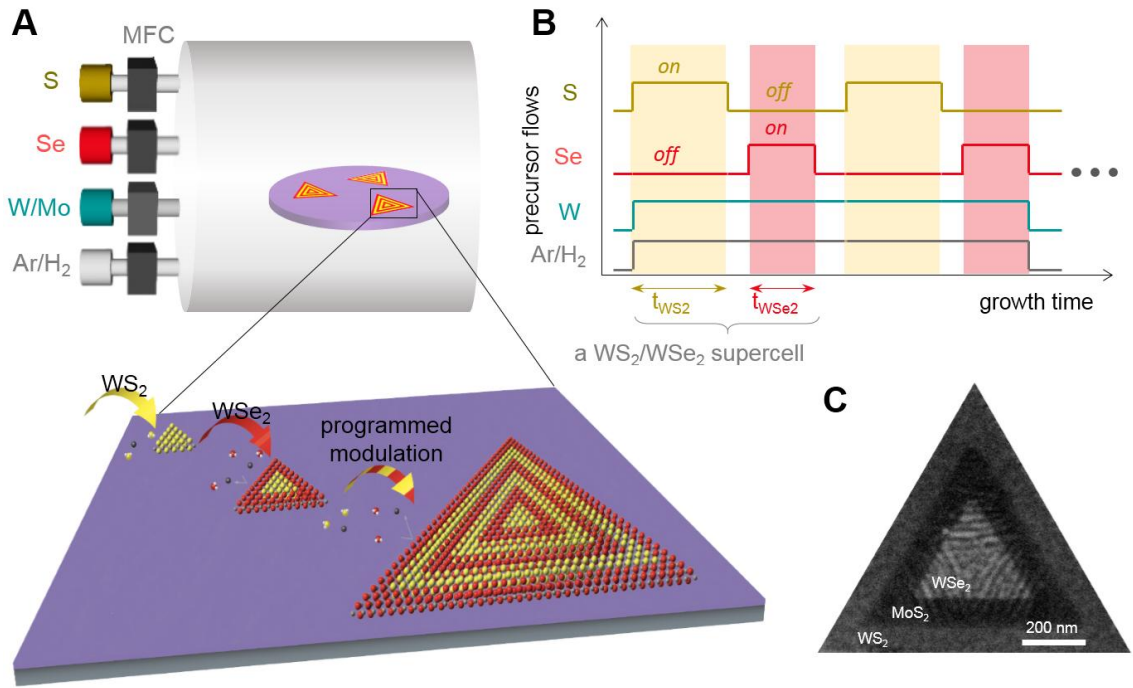

**Fig. S1. Modulated-MOCVD process.** (**A**) Top, schematic of modulated-MOCVD reactor with individual MFCs that precisely control the flow of all precursors. Bottom, schematic of the programmed epitaxial growth of a $WS_2/WSe_2$ superlattice. (**B**) Time sequence of the modulated superlattice growth where the growth time for a supercell {$WS_2$, $WSe_2$} is {$t_{WS2}$, $t_{WSe2}$}. (**C**) SEM image of a coherent $WSe_2/MoS_2/WS_2$ heterostructure.

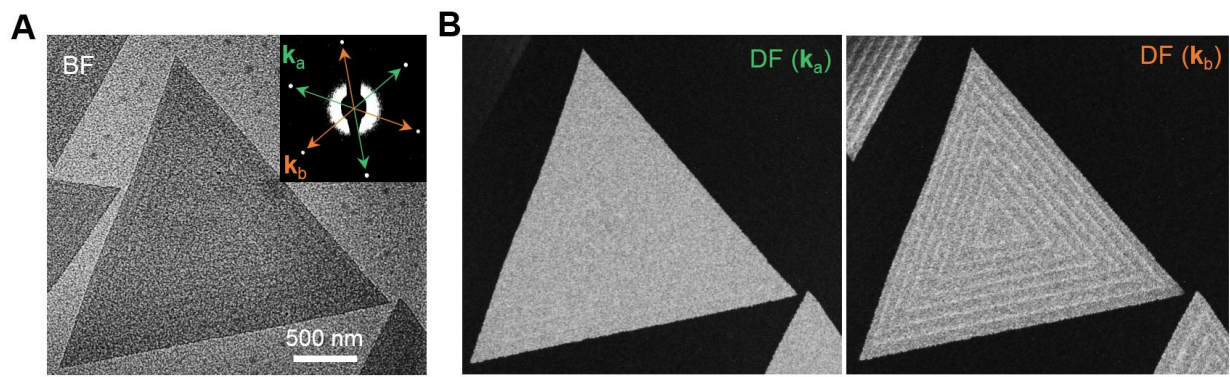

**Fig. S2. TEM images of a $WS_2/WSe_2$ superlattice.** (**A**) Bright-field image of superlattice {50 nm, 40 nm}. Inset: diffraction patterns with the first order spots divided into two families of $k_a$ and $k_b$. Scale bar, 500 nm. (**B**) Dark-field TEM images acquired by collecting from $k_a$ and $k_b$, respectively. The uniform contrast shown in DF ($k_a$) suggests that $k_a$ points towards the W-zigzag direction (*36*).

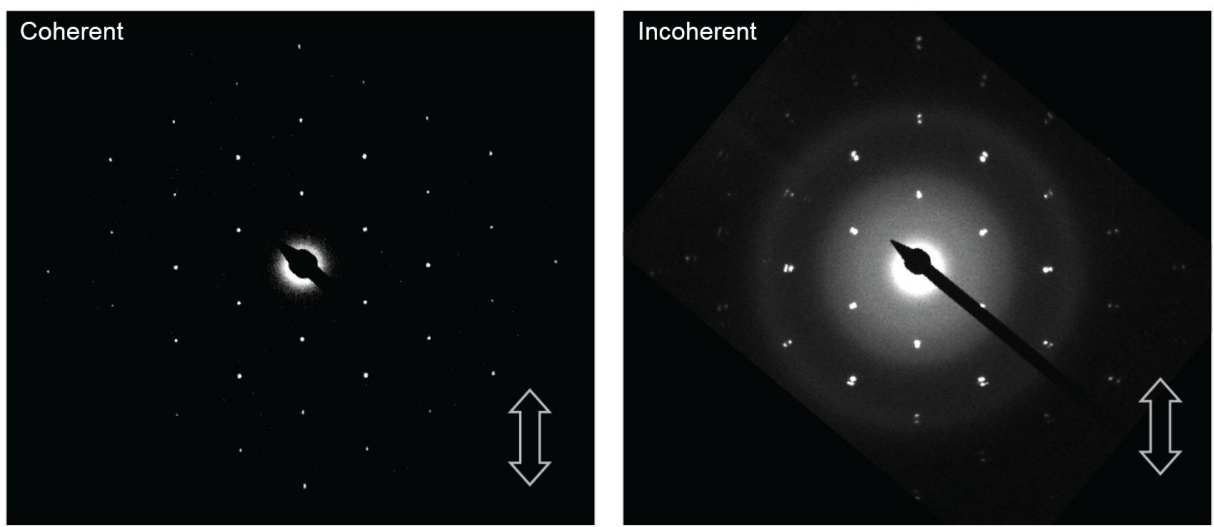

**Fig. S3. SAED patterns of WS$_2$/WSe$_2$ heterostructures.** Left, enlarged SAED pattern shown in Fig. 2C. Right, the full SAED pattern of an incoherent heterostructure from which Fig. 2E is taken.

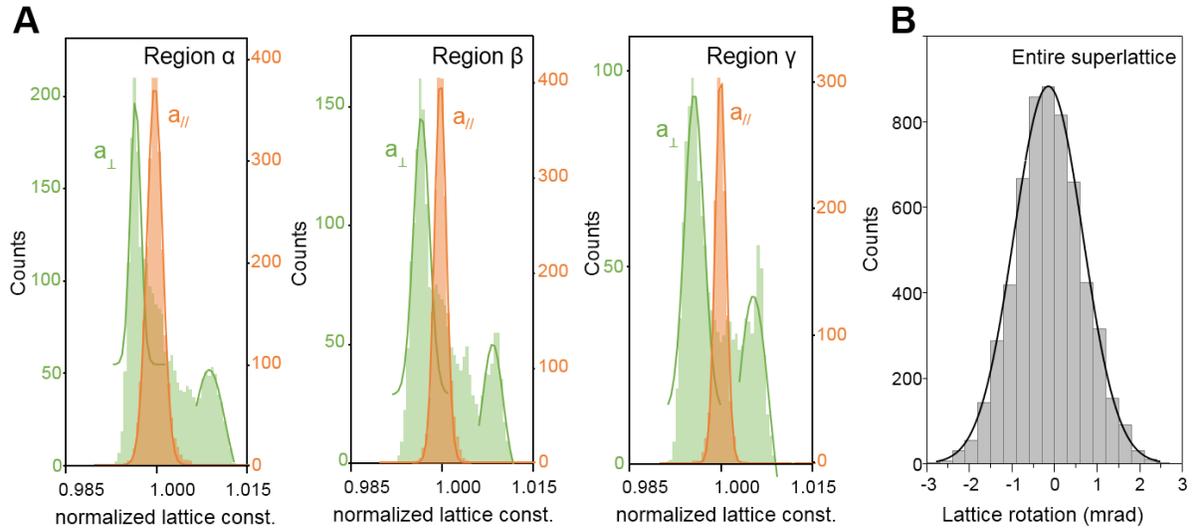

**Fig. S4. Histograms of lattice parameters.** (**A**) Histograms of $a_{//}$ and $a_\perp$ for region α (two $a_\perp$ peaks centered at 0.996 and 1.008), β (two $a_\perp$ peaks centered at 0.996 and 1.008), and γ (two $a_\perp$ peaks centered at 0.996 and 1.005). (**C**) Histogram of lattice rotation of the entire superlattice showing a standard deviation of 0.8 mrad.

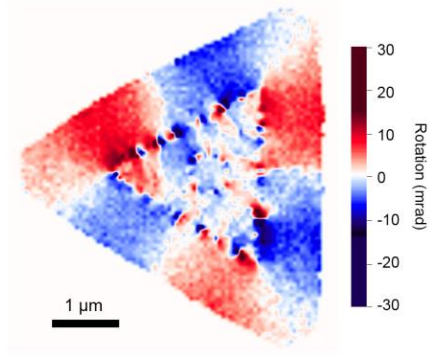

**Fig. S5. Lattice rotation map of an incoherent $WS_2/WSe_2$ heterostructure.** An incoherent $WS_2/WSe_2$ heterostructure showing arrays of dislocations at the heterointerfaces. Scale bar, 1 μm.

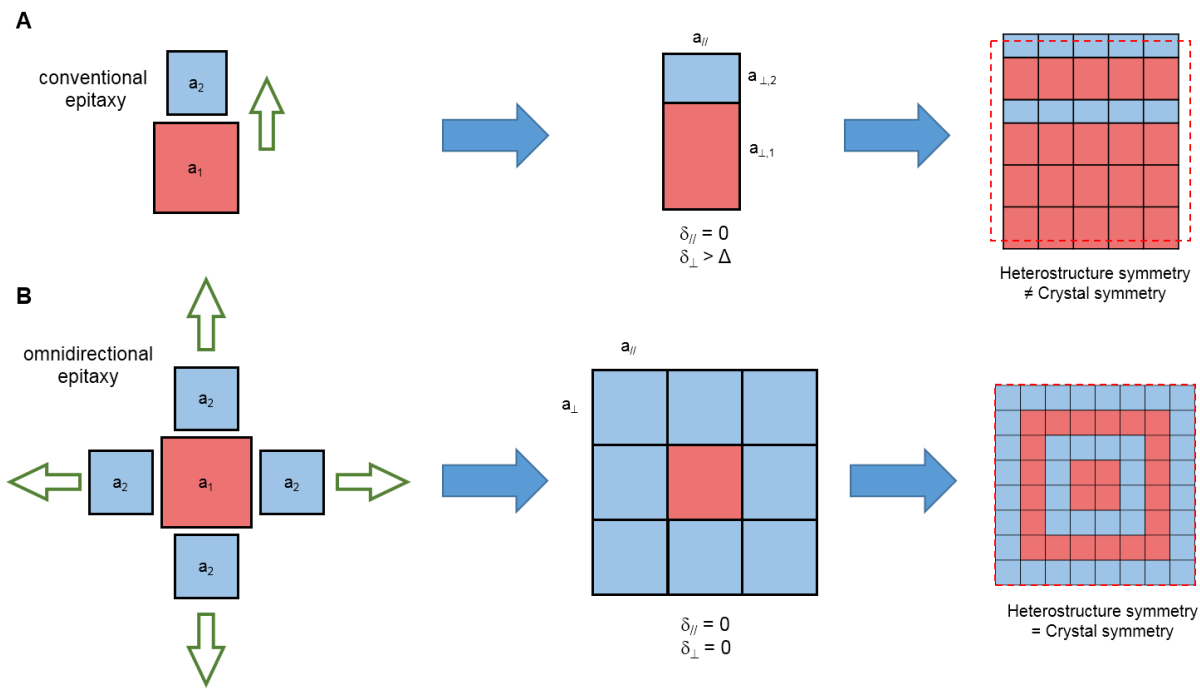

**Fig. S6. Schematic of conventional and omnidirectional epitaxy.** (**A**) Schematic of conventional unidirectional epitaxy and the resulting heterostructure with different symmetry from that of the original crystal. (**B**) Schematic of omnidirectional coherent epitaxy and the resulting heterostructure maintaining the same symmetry of the original crystal.

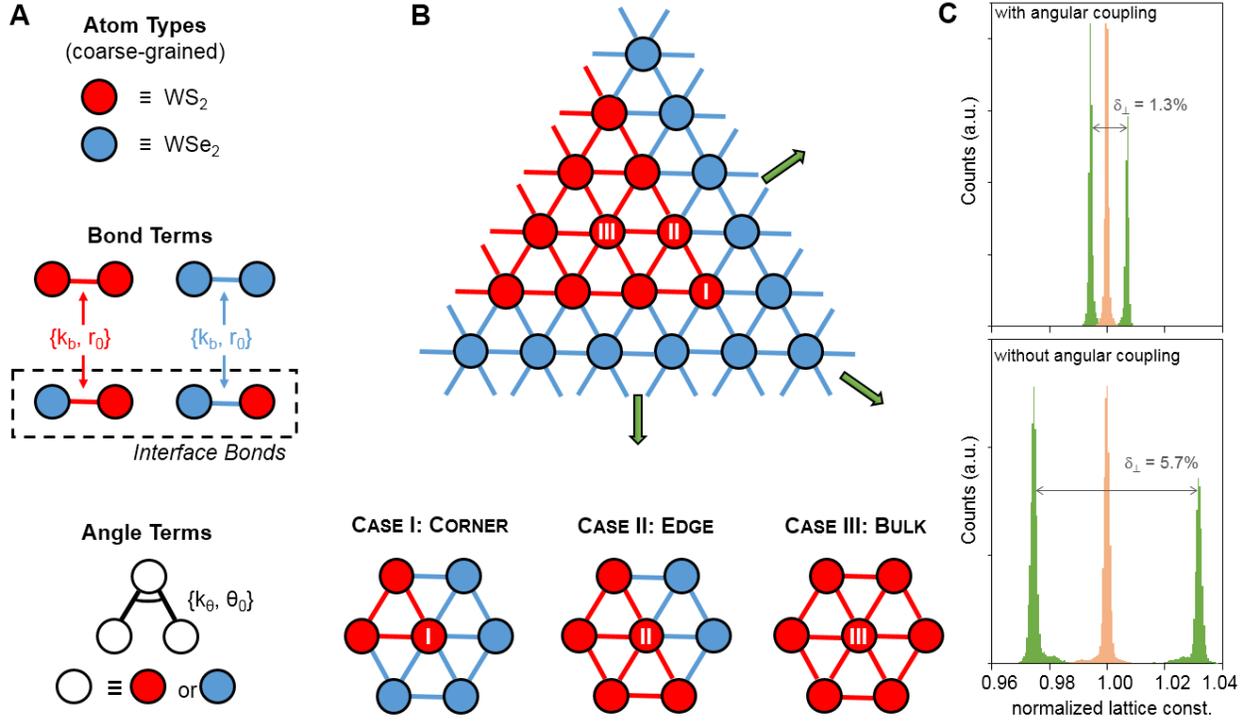

**Fig. S7. Coarse-grained simulation of the WS$_2$/WSe$_2$ coherent superlattice.** (**A**) Red and blue circles depict the atom types in this system: "dressed W atoms" that can either be [WS$_2$] or [WSe$_2$] units. There are four types of bond terms, depending on the dressed atom types and the interface conditions: [WS$_2$]—[WS$_2$], [WSe$_2$]—[WSe$_2$], [WS$_2$]—[WSe$_2$] (growth from [WS$_2$] to [WSe$_2$]) and [WSe$_2$]—[WS$_2$] (growth from [WSe$_2$] to [WS$_2$]). For the angular interactions, each of the three dressed atoms that form the angle can either be [WS$_2$] or [WSe$_2$]. (**B**) Three different environments for dressed atoms on an interior corner of the WS$_2$/WSe$_2$ superlattice. Green arrows denote the growth direction in the TMD superlattice. (**C**) Theoretical simulation results for the a$_{//}$ and a$_\perp$ distributions. The addition of angular coupling terms significantly reduces $\delta_\perp$ from 5.7 % to 1.3 %, in excellent agreement with the experimental observation of coherent epitaxy in both the parallel and perpendicular directions.

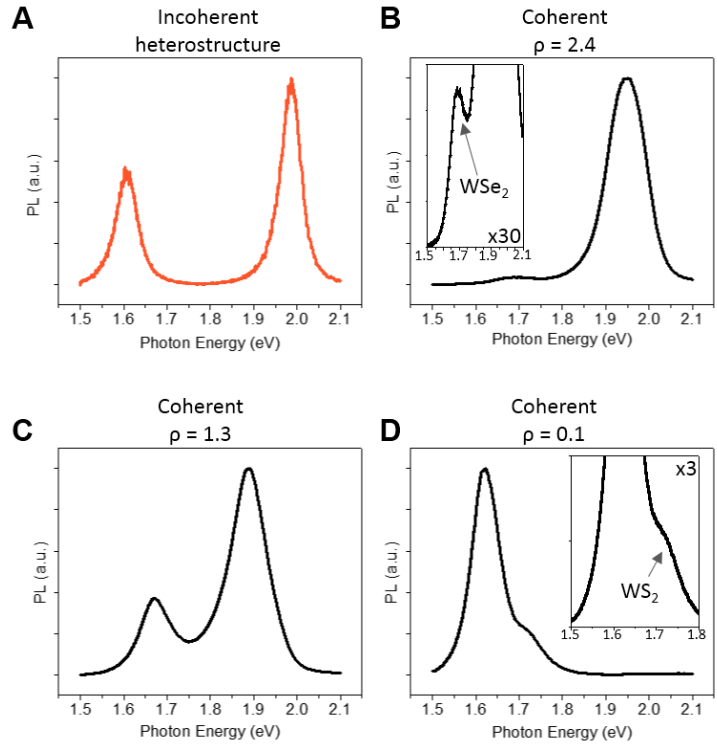

**Fig. S8. PL spectra of WS$_2$/WSe$_2$ heterostructures.** PL spectra of an incoherent WS$_2$/WSe$_2$ heterostructure (**A**) and three coherent WS$_2$/WSe$_2$ superlattices with $\rho$ = 2.4 (**B**), 1.3 (**C**) and 0.1 (**D**). Insets show enlarged spectra for clarity.

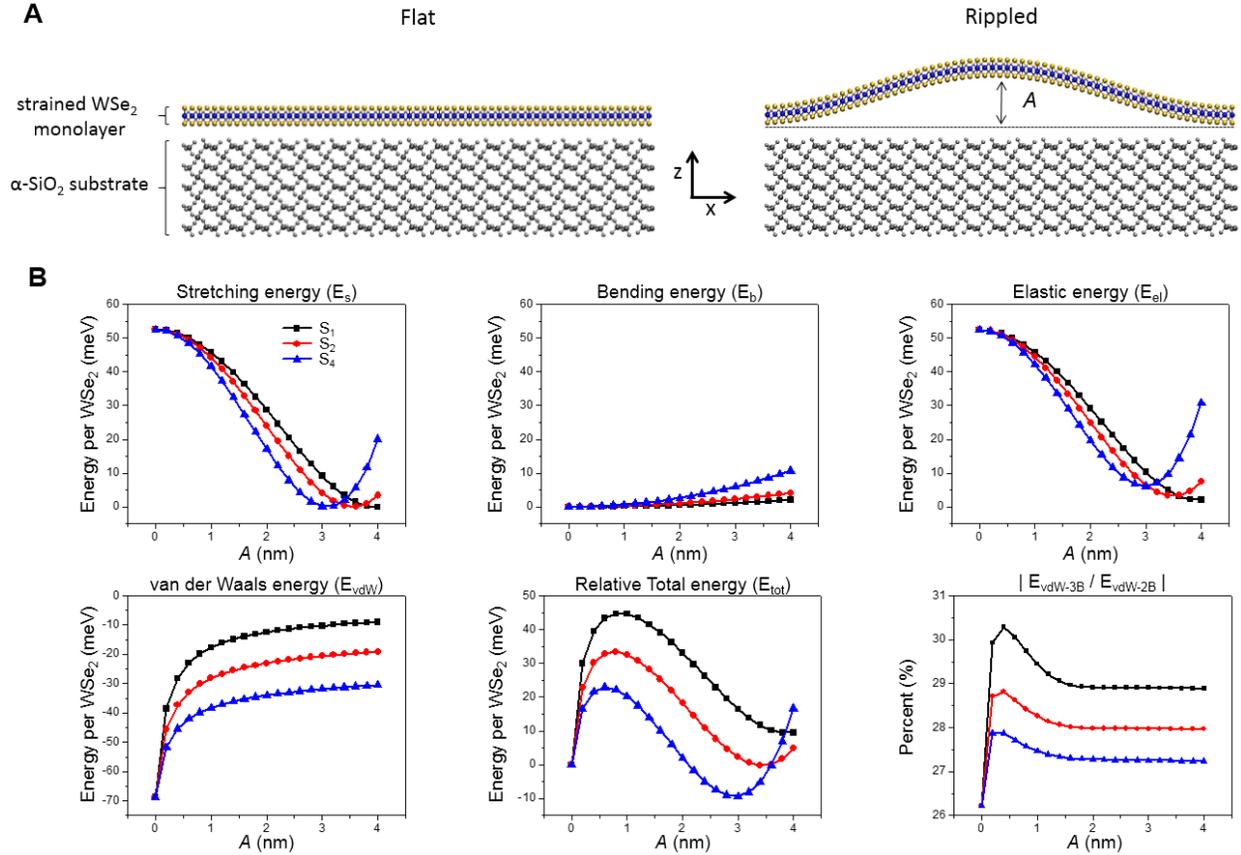

**Fig. S9. Energetics of rippled WSe$_2$.** (**A**) Schematic of the model for the flat-rippled configuration space in a strained WSe$_2$ monolayer grown on a SiO$_2$ substrate. (**B**) Stretching energy (E$_s$, upper left), bending energy (E$_b$, upper middle), elastic energy (E$_{el}$ = E$_s$ + E$_b$, upper right), interlayer van der Waals binding energy (E$_{vdW}$, lower left), and relative total energy (E$_{tot}$ = E$_{el}$ + E$_{vdW}$, lower middle) per WSe$_2$ as a function of the ripple height (*A*) for WSe$_2$ ripple shapes generated using the sinusoidal S$_1$, S$_2$, and S$_4$ transformations (see supplementary text for details). Lower right panel: ratio of three- to two-body interlayer vdW binding energy contributions, |E$_{vdW-3B}$/E$_{vdW-2B}$|, as a function of *A* for the various ripple shapes.

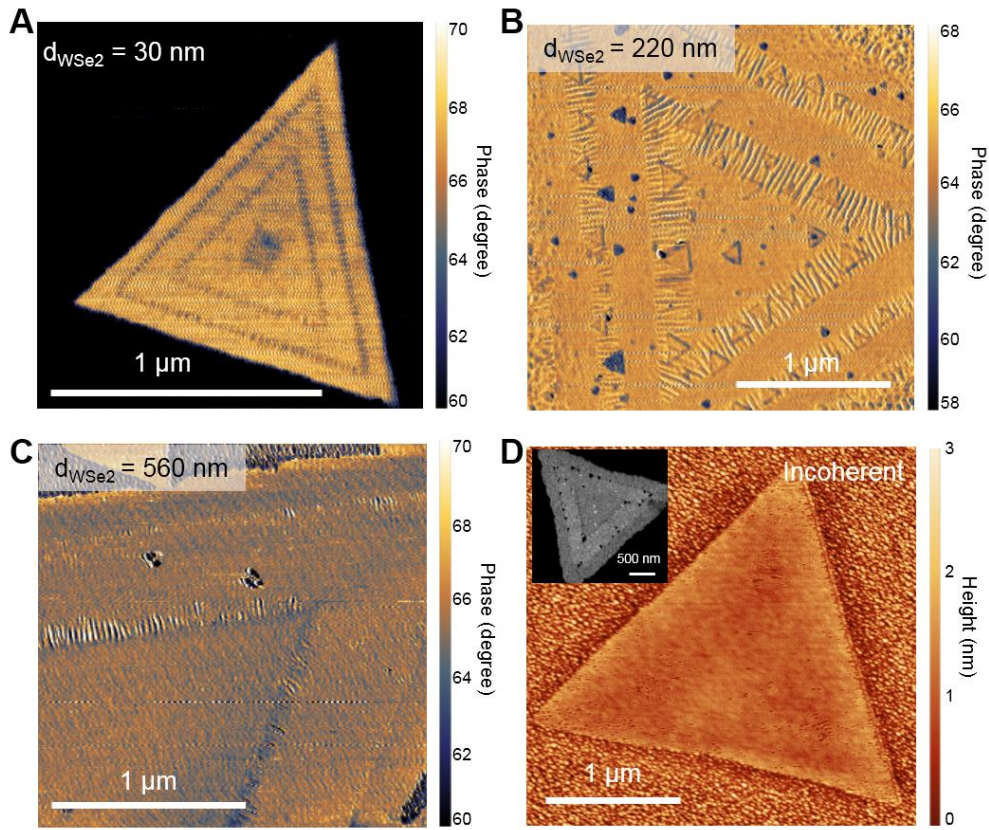

**Fig. S10. AFM images of WS₂/WSe₂ heterostructures.** (**A**-**C**) AFM phase images of WS$_2$/WSe$_2$ heterostructures with different d$_{WSe2}$ of 30, 220, and 560 nm, respectively. Ripples are not continuous across the entire WSe$_2$ region in the superlattice with d$_{WSe2}$ of 560 nm. (**D**) AFM height image of an incoherent WS$_2$/WSe$_2$ heterostructures showing no out-of-plane ripples. Inset: DF-TEM image of a similar sample showing the WS$_2$/WSe$_2$/WS$_2$ composition.

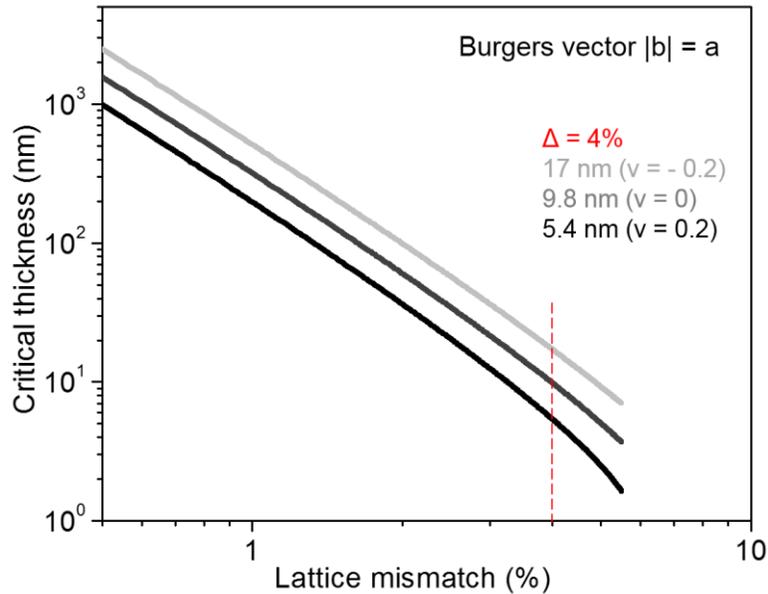

**Fig. S11. Critical thickness.** Critical thickness as a function of lattice mismatches, for $\nu = -0.2$ (light gray), 0 (gray), and 0.2 (black), based on the People-Bean model. A critical thickness of 17 nm ($\nu = -0.2$), 9.8 nm ($\nu = 0$), and 5.4 nm ($\nu = 0.2$) is estimated for the $WS_2/WSe_2$ epitaxial system with lattice mismatch of 4%.

**References for Supplementary Materials:**